\documentclass{jps-cp}

\usepackage{amsmath}
\usepackage[dvipsnames]{xcolor}
\usepackage{upgreek}
\title{Resistive Read-out in Thin Silicon Sensors with Internal Gain}

\author{N. \textsc{Cartiglia}$^{1}$,  F.  \textsc{Moscatelli}$^{4,6}$,
R.   \textsc{Arcidiacono}$^{1,2}$,
P.  \textsc{Asenov}$^{5,6}$,
M.  \textsc{Costa}$^{1,3}$,
T. \textsc{Croci}$^{6}$,
M.  \textsc{Ferrero}$^{1}$,
A. \textsc{Fondacci}$^{5,6}$, 
L.  \textsc{Lanteri}$^{1,3}$,
L.  \textsc{Menzio}$^{1,3}$,
  A. \textsc{Morozzi}$^{6}$,
R.  \textsc{Mulargia}$^{1,3}$,
 D.\textsc{Passeri}$^{5,6}$,
F.  \textsc{Siviero}$^{1}$,
V.  \textsc{Sola}$^{1,3}$,
M. \textsc{Tornago}$^{1,3}$
}

\inst{
$^{1}$  INFN, Torino, Italy,
$^{2}$ Universita del Piemonte orientale, Italy
$^{3}$ Universita di Torino, Italy,
$^{4}$ CNR-IOM, Perugia, Italy, 
$^{5}$ Universita di Perugia, Italy.
$^{6}$ INFN Perugia, Italy.
}

\email{cartiglia@to.infn.it}

\abst{Two design innovations,  low-gain avalanche (Low-Gain Avalance Diode, LGAD) and resistive read-out (Resistive Silicon Detector, RSD), have brought strong performance improvements to silicon sensors. Large signals, due to the added gain mechanism,  lead to improved temporal precision, while charge sharing, introduced by resistive read-out,  allows for achieving excellent spatial resolution even with large pixels. LGAD- and RSD- based silicon sensors are now adopted, or considered, in several future experiments and are the basis for almost every next 4D-trackers. New results obtained with sensors belonging to the second FBK production of RSD (RSD2) demonstrate how a combined resolution of 30 ps and 30 \microns can be obtained with pixels as large as  $1 \times 1 $ mm$^2$.}

\newcommand{\micronsqs}{$\upmu$m$^2$~}
\newcommand{\micronsq}{$\upmu$m$^2$}
\newcommand{\microns}{$\upmu$m~}
\newcommand{\micron}{$\upmu$m}
\newcommand{\shits}{$\sigma_{hit\; pos}$~}
\newcommand{\smss}{$\sigma_{MS}$~}
\newcommand{\shit}{$\sigma_{hit\; pos}$}
\newcommand{\sms}{$\sigma_{MS}$}

\kword{resistive read-out, silicon sensors, low-gain avalanche diode}

\begin{document}

\maketitle

\section{Introduction}

In the past few years, the performance capabilities of silicon sensors in terms of combined spatial and temporal resolutions have improved significantly. This evolution is due to the introduction of two innovations in the design of silicon sensors: (i) Low-Gain Avalanche Diode and (ii) Resistive Read-out.

The Low-Gain Avalanche Diode~\cite{FERNANDEZMARTINEZ201198, LGAD1} design has been first introduced to compensate for the loss of signal due to charge trapping in irradiated sensors. However, this design found its main application in the field of precision timing, with the introduction of  Ultra-Fast Silicon Detector (UFSD)~\cite{CARTIGLIA2015141}. This R\&D has spurred a strong evolution in the field of accurate timing using silicon detectors.

Resistive read-out was first proposed to achieve an LGAD design with 100\% fill factor~\cite{TREDI2015N} (the so-called AC-LGAD). Then, it was subsequently recognized to lead to excellent spatial precision with a concurrent reduction of the number of read-out electrodes~\cite{Siviero_2021, tornago2020resistive}. Sensors based on resistive read-out are called RSDs (Resistive Silicon Detectors). In the productions completed so far ~\cite{2019Giacomini_JINST,Heller_2022,Mandurrino:2021cuq}, RSDs have an AC-coupled design; recently, the design has been extended to DC-coupled read-out~\cite{MenzioVCI2022} (DC-RSD).  The key feature of RSDs is that the signal is shared among the electrodes near the hit point analogously to a current divider: each pad $i$ sees a fraction $I_i$ of the total signal $I_o$ that depends on the impedance $Z_j$ between the impact point and the pads, $ I_i = I_o (1/Z_i)/\sum_1^n (1/Z_j)$. 
 
\begin{figure}[htb]
\begin{center}
\includegraphics[width=0.8\textwidth]{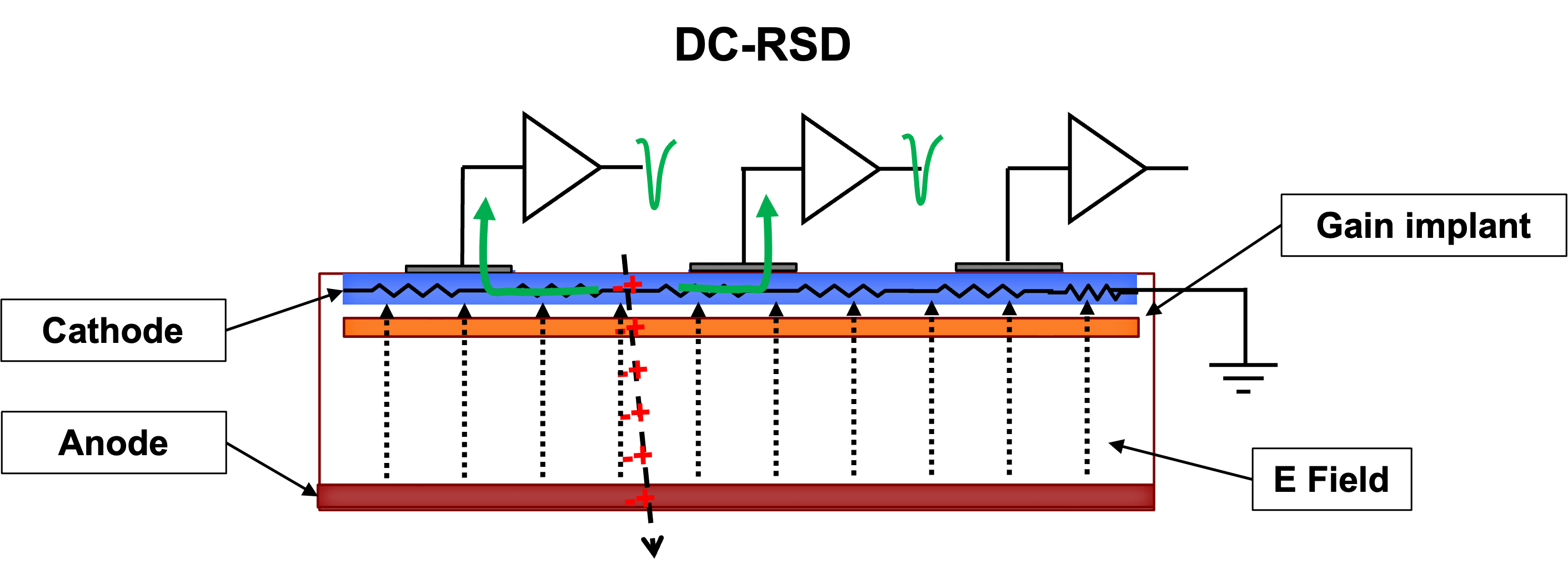}
\caption{Sketch of a resistive silicon detector with DC read-out (DC-RSD).}. 
\label{fig:rsd}
\end{center}
\end{figure}

 Figure~\ref{fig:rsd} shows the sketch of a silicon sensor that incorporates both the LGAD and RSD innovations. The design is based on an n-in-p sensor, has a continuous gain implant just underneath the cathode, and the cathode is resistive to ensure electrode isolation and signal sharing. The presence of the gain implant, the signature feature of the LGAD design, creates a high electric field in the volume underneath the $n^+$ resistive layer and leads to signal multiplication. In this sketch, the metal electrodes are directly implanted in the resistive $n^+$ sheet to ensure a DC coupling between the sensor and the electronics (DC-RSD).   The RSD design has uniform electric and weighting fields over the whole sensor volume, a requisite for good temporal resolution. 
 
 \section{Future Silicon Trackers}
 
According to the ECFA roadmap~\cite{Detector:2784893}, the next generation of large silicon trackers will be deployed at either circular (FCC) or linear (ILC, CLIC) lepton colliders.  The physics aims of these future lepton colliders, revolving around very precise flavor physics,  can only be achieved with silicon trackers with very small impact parameter resolutions. Table~\ref{tab:param} reports a few important parameters of future silicon vertex trackers at e$^+$e$^-$ colliders.

\begin{table}[tbh]
\begin{center}
\caption{Parameters of future silicon vertex trackers at e$^+$e$^-$ colliders.}
\label{tab:param}
\begin{tabular}{ | c | c  | c  |c  |} \hline
Facility: & FCC-ee & ILC & CLIC \\ \hline
\shits [\micron] & $\sim 5 $ & $<3$ & $<3$\\ \hline
Thickness [\microns of Si] & $\sim 100$ & $\sim 100$ & $\sim 100$ \\ \hline
Hit rate [$10^6$/s/cm$^2]$ & $\sim 20 $ & $\sim 0.2 $ & $\sim 1 $  \\ \hline
Power dissipation [W/cm$^2$] & 0.1 -0.2 & 0.1 & 0.1 \\ \hline
Pixel size [\micronsq] & $25 \times 25$ & $25 \times 25$   & $25 \times 25$     \\ \hline
\end{tabular}
\end{center}
\end{table}

The trackers need to have a superb position resolution (less than 5 \micron), be very thin, sustain a high hit particle rate, and use very little power.  Two quantities need to be minimized to achieve these goals: (i) the single hit resolution \shit, and (ii) the multiple scattering resolution \sms. 
The two terms \shits and \smss are linked to each other and to the type of read-out architecture. The pixel size determines the spatial resolution if a tracker employs single-pixel read-out. Only very small pixels (25x25 \micronsq) achieve the required precision of 5 \micron, and it is practically impossible to reach better resolutions.  On the other hand, if a tracker uses the traditional design for charge sharing to improve \shit, the sensor needs to be quite thick (at least 200 \micron), leading to a large \smss contribution. The very mechanism that optimizes \shits is detrimental to \sms: thick sensors, necessary for signal sharing, cause significant multiple scattering and deteriorate the overall accuracy of the tracker system. 
Multiple scattering is further increased by the cooling infrastructures: for this reason, the levels of power consumption reported in Table~\ref{tab:param} are such that air cooling can be employed. This request is particularly difficult to achieve when using single pixel read-out as the power used by millions of pixels is large.  Although research and development in silicon detectors is very active in many fields, currently, no design can achieve the performance listed in Table~\ref{tab:param}~\cite{ CARTIGLIA2022167228}.

\section{A Tracker Based on Resistive Read-out in Thin Silicon Sensors with Internal Gain}
According to our present R\&D studies, the key to meeting the demand of the next generation of lepton colliders is to use thin silicon sensors that combine resistive read-out and internal gain.  

In the present silicon tracker paradigm, the targeted position resolution determines the pixel size. These systems reach an excellent spatial resolution using millions of pixels and amplifiers that, at any given time, are mostly empty; in many situations, less than 0.1 \% of pixels see a signal. In a much more efficient design, the density of particles determines the pixel size. For example, the pixel size should be such that less than a few per mille of pixels are hit by 2 particles at the same time. Resistive read-out allows reaching excellent position resolution while using large pixels, so the pixel size is determined by occupancy and not by the needed position resolution.  The presence of internal gain boosts the signal and allows using thin sensors.

\begin{figure}[htb]
\begin{center}
\includegraphics[width=0.8\textwidth]{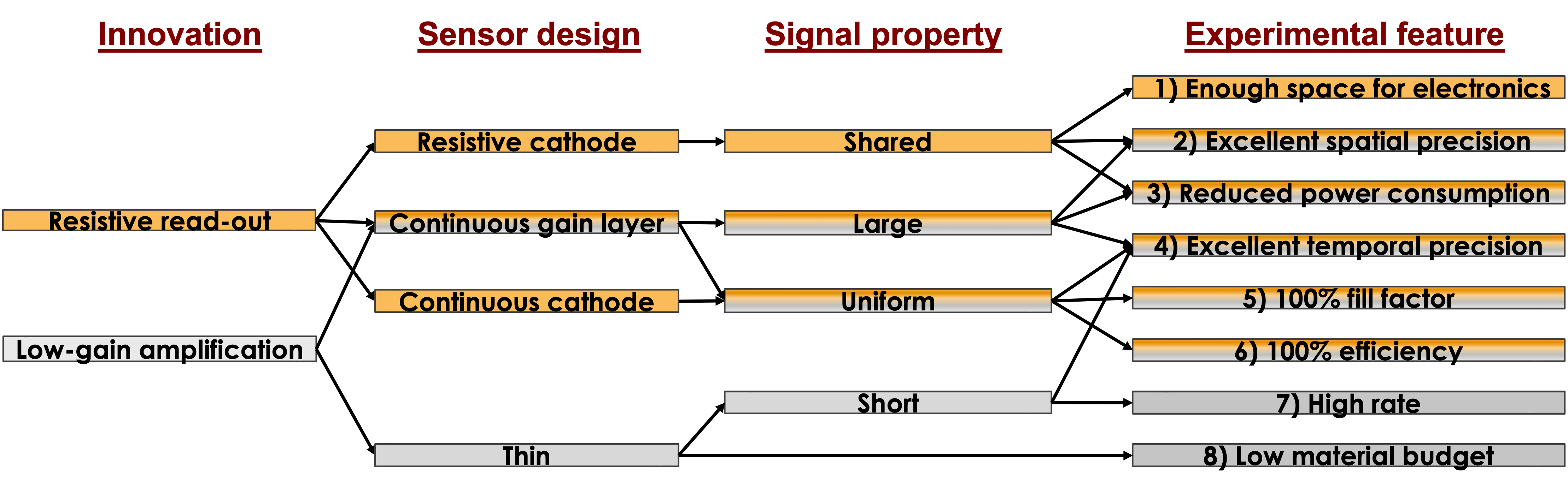}
\caption{The combination of resistive read-out and low-gain amplification leads to the experimental features required by the next generation of experiments.} 
\label{fig:Signal}
\end{center}
\end{figure}

Figure~\ref{fig:Signal}   illustrates why the combination of resistive read-out and low-gain amplification is so powerful.
With this design, the signals are shared among a few pads, have large amplitudes, are uniform over the sensor surface, and are short.

The experimental features arising from the design innovations can be summarised as:
\begin{enumerate}
\item Sharing allows using large pixels, i.e., having enough space for the read-out electronics.
\item Sharing combined with internal amplification achieves excellent spatial precision.
\item Sharing (fewer pixels) combined with internal amplification (less need for amplification in the electronics)
reduces power consumption.
\item Large and short signals combined with a uniform response yield excellent temporal precision.
\item Large signals and a uniform response yield 100\% fill factor.
\item Large signals and a uniform response yield 100\% efficiency.
\item Short signals make it possible to work at a high rate.
\item Thin sensors (and thin front-end electronics) enable a low material budget.
\end{enumerate}

\subsection{Sensor Simulation}
\begin{figure}[htb]
\begin{center}
\includegraphics[width=0.8\textwidth]{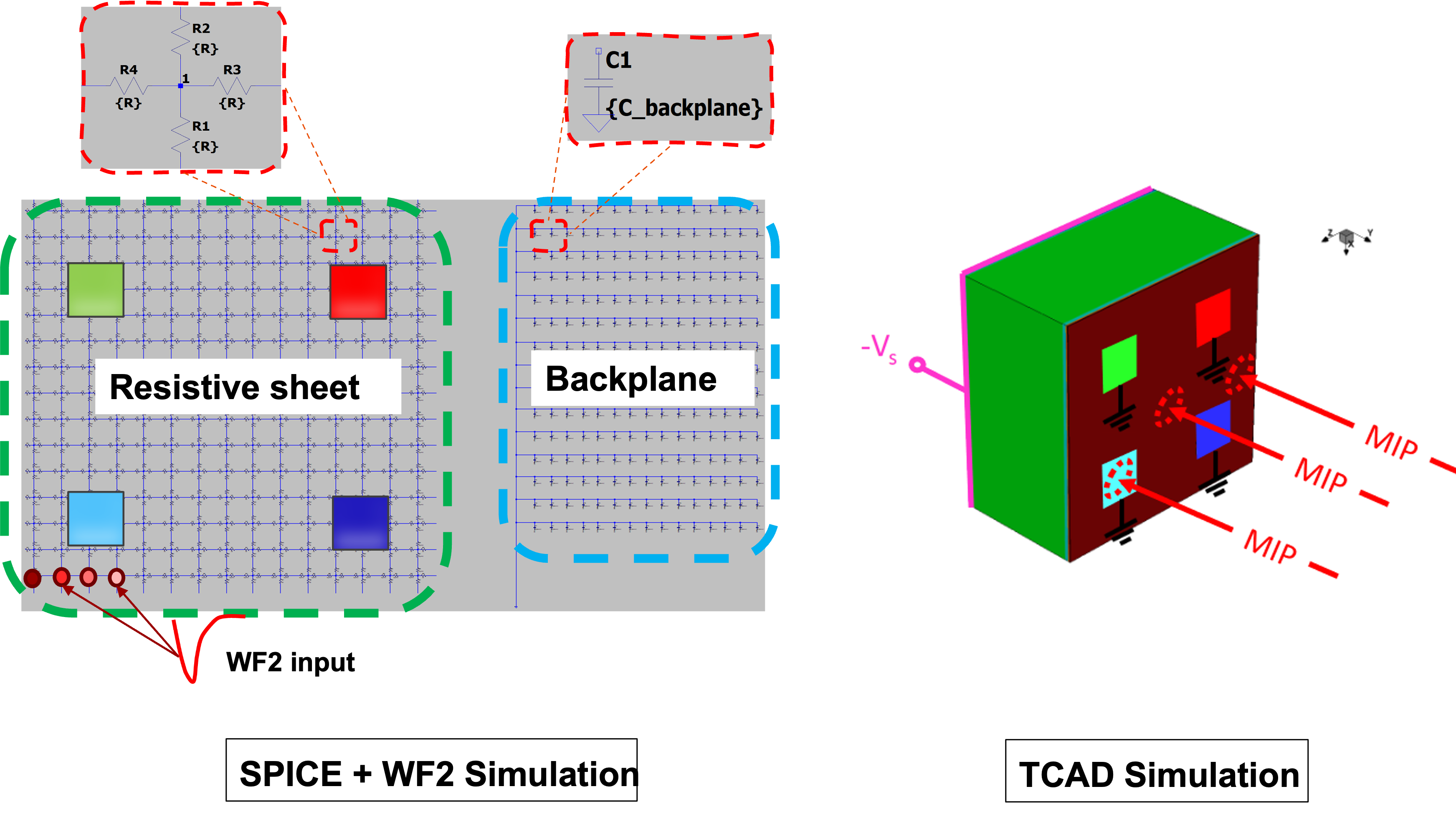}
\caption{A graphical representation of the mixed-mode approach for the simulation of RSD. Left pane: the building block of the SPICE-based RSD model. The simulation package Weightfield2 (WF2) is used to generate the input signals. Right pane: the 3D TCAD volume, with 4 pads. } 
\label{fig:Sim}
\end{center}
\end{figure}

The simulation of RSD presents several unique challenges linked to the complex nature of its design and to the large pixel size. The defining feature of RSD, built-in charge sharing over distances that can be as large as a millimeter, represents a formidable challenge for TCAD, the standard simulation tool.  A single 3D TCAD simulation of an  RSD pixel $100 \times 100$ \micronsqs  takes about 12 hours. Since the simulation time is an increasing function of the simulated volume, the time needed to perform a 3D simulation of a $1 \times  1$ mm$^2$ pixel is too long to be used in an R\&D phase, where many different options need to be tested. It is, therefore, impossible to approach the simulation of RSDs using the standard TCAD method.  To circumvent this problem, a mixed-mode approach to simulation was developed~\cite{TCroci}, shown graphically in Figure~\ref{fig:Sim}.

In the first phase, the properties of RSD are simulated using a SPICE-based analog electronic circuit software. In this step, the key elements of an RSD are represented by electrical components (left side of  Figure~\ref{fig:Sim}). The resistive plane is modeled by a network of resistors, the sensor bulk by capacitors, the metal pads by areas with zero-ohm resistance, and the front-end electronics by resistors that approximate the read-out input impedance. In this framework, signal sharing is studied by injecting a current stimulus in pre-determined positions on the resistive plane and measuring the current collected in each pad. With this approach, each simulation takes about 1 minute. A very important aspect of this simulation approach is the shape of the current stimulus. Since the propagation of a signal on an RC network depends on its frequency components, the signals used in this study need to have the appropriate shape. To this end, the Weightfield2 (WF2)  simulation package~\cite{CENNA201514}, has been used. WF2 is a software program specifically developed to simulate signal formation in silicon sensors with or without gain, and it can provide accurate predictions of the induced current signal in RSD.  

The outcome of this first phase is used in 3D TCAD simulations to determine the actual sensor design. The most promising values used in SPICE are converted into actual design parameters. For example, the n$^+$ sheet resistance is obtained with appropriate n$^+$  electrode doping and profile combinations. Once the construction parameters of RSD are finalized, a set of 3D TCAD simulations of small-sized pixels are performed to cross-check the SPICE-based approach predictions. The right side of Figure~\ref{fig:Sim} shows the volume of an RSD device, with an area of 100 $\times$ 100 \micronsq, in a 3D TCAD simulation with four read-out pads. 

\subsection{The design of the electronics}
\begin{figure}[htb]
\begin{center}
\includegraphics[width=0.9\textwidth]{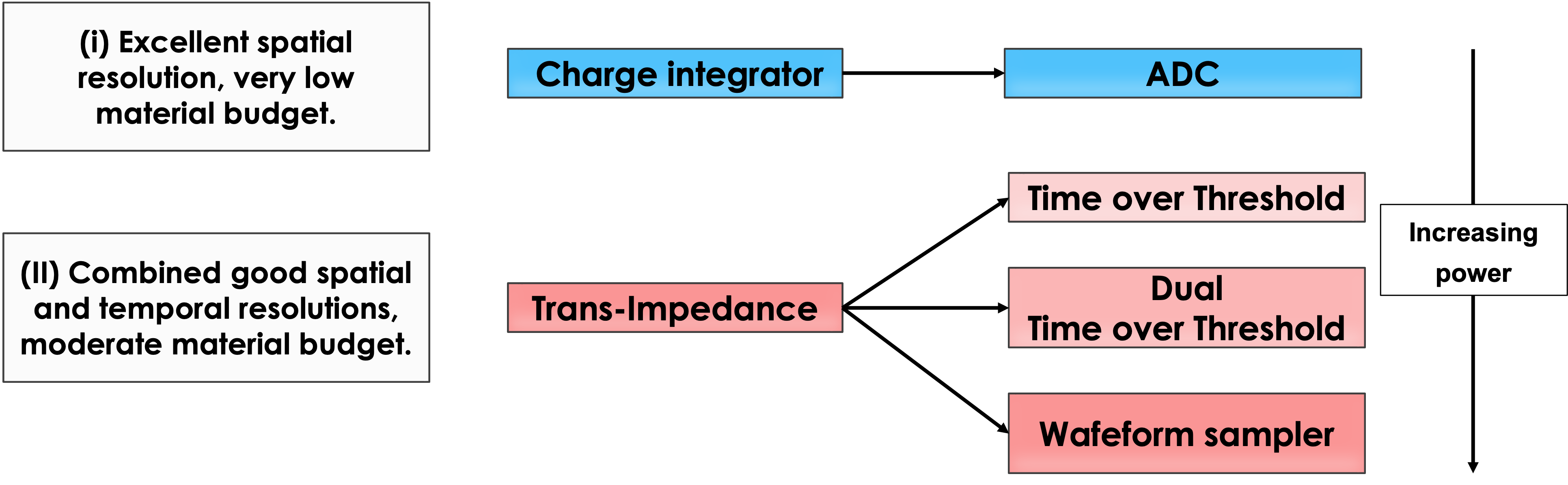}
\caption{First and second stage architectures for the two options considered in the text. } 
\label{fig:Elec}
\end{center}
\end{figure}

The signals generated by particles in the RSD sensor define how the first amplification stage should be designed.
The electronic design differs considerably depending on the tracker goals.  \begin{itemize}
\item {\bf Excellent spatial resolution, stringent requirements on material budget.} For this configuration, the most important aspect is to precisely measure the charge on each pad and keep the power consumption very low. Therefore, the most promising front-end configuration is a charge integrator followed by an ADC. This design needs very low power, and it can match the requirement of less than 80-100 mW/cm$^2$, the maximum power budget where air-cooling can be used. 
\item {\bf Combined good spatial and temporal resolutions, moderate requirements on material budget.}For this configuration, a time tagging circuit is necessary, which leads to higher power consumption.  Depending on the spatial precision required, the signal can be sampled once using the standard Time-over-Threshold (ToT) approach, twice with a dual ToT system, or multiple times using a waveform sampler (WFS). The choice among ToT, dual ToT, and WFS depends upon the power available and the required precision. 
\end{itemize}
Figure~\ref{fig:Elec} schematically shows the options for the two cases.

In RSDs, each pad sees a modified version of the original signal. During the propagation on the n$^+$ resistive surface, the signal becomes smaller, wider, with slower leading and trailing edges, and is delayed. Each of these aspects is a valuable piece of information that should be used in the reconstruction of the hit position and time. For this reason, the finer the signal sampling, the better the hit location can be determined. The ToT solution provides only a limited subset of information to the reconstruction stage and might lead to a degradation of performance. Double ToT considerably increases the information available to the reconstruction as it provides the signal derivative of the signal leading and trailing edges. A waveform sampler  (WFS) is the option that provides the most information, however, at the price of a steep increase in power consumption. 

\subsection{The Design of a New  Silicon Tracker}

\begin{figure}[htb]
\begin{center}
\includegraphics[width=0.8\textwidth]{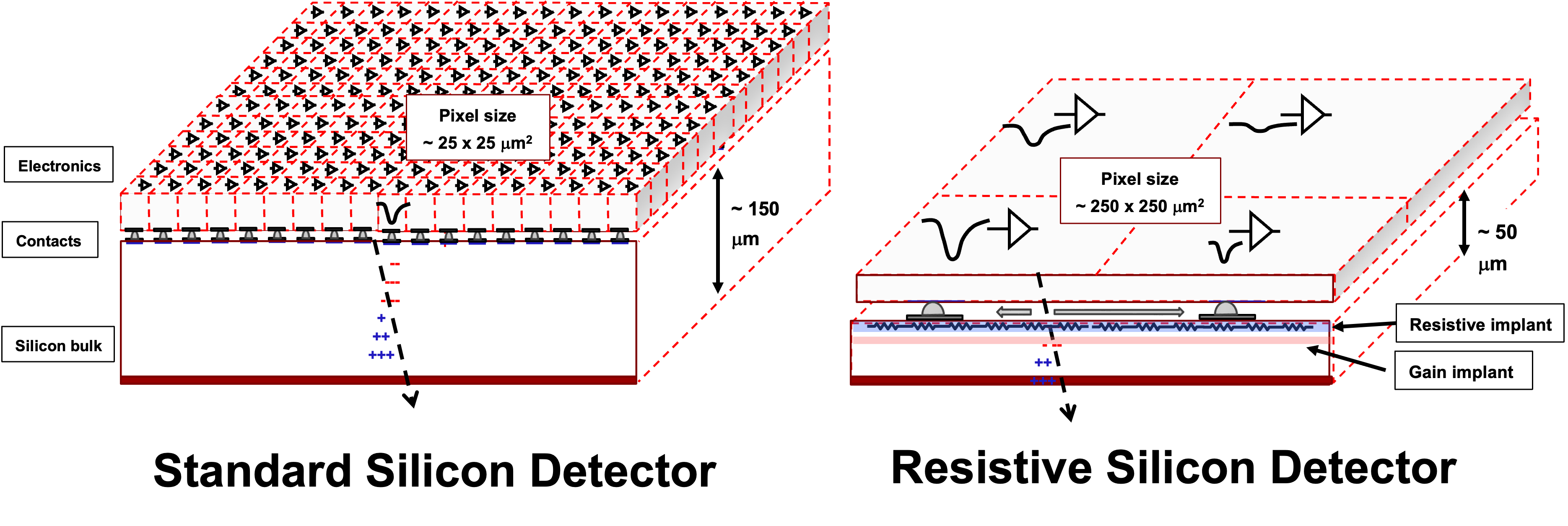}
\caption{Sketch of a present (left) and RSD-based (right) silicon detector.  The two detector designs yield the same spatial precision. However, to cover an area of 600 x 600 \micronsqs the standard detector uses about 575 pixels while the RSD uses 4 pixels.} 
\label{fig:RSD}
\end{center}
\end{figure}

Figure~\ref{fig:RSD} shows the transformation brought about by the combination of resistive read-out and internal gain. In present systems (left pane), sensors are made of many independent p-n diodes, each with its own electronics. A minimum thickness of about 150 \microns is needed to ensure good efficiency. In the RSD design (right pane), there is a single p-n diode with a p-doped bulk and an n-doped resistive cathode. The signal is boosted by built-in amplification, generated by the high field created by an extra p-doped layer, and it is shared among contiguous pixels. The sensor thickness is 30-50 \micron. The large sizes of the cathode and anode are also instrumental in providing a very uniform weighting field, and uniform charge carriers drift velocities, two key features to achieving 100\% signal uniformity, 100\% efficiency, and excellent temporal resolution.

\section{The Second FBK RSD production}
\begin{figure}[htb]
\begin{center}
\includegraphics[width=0.6\textwidth]{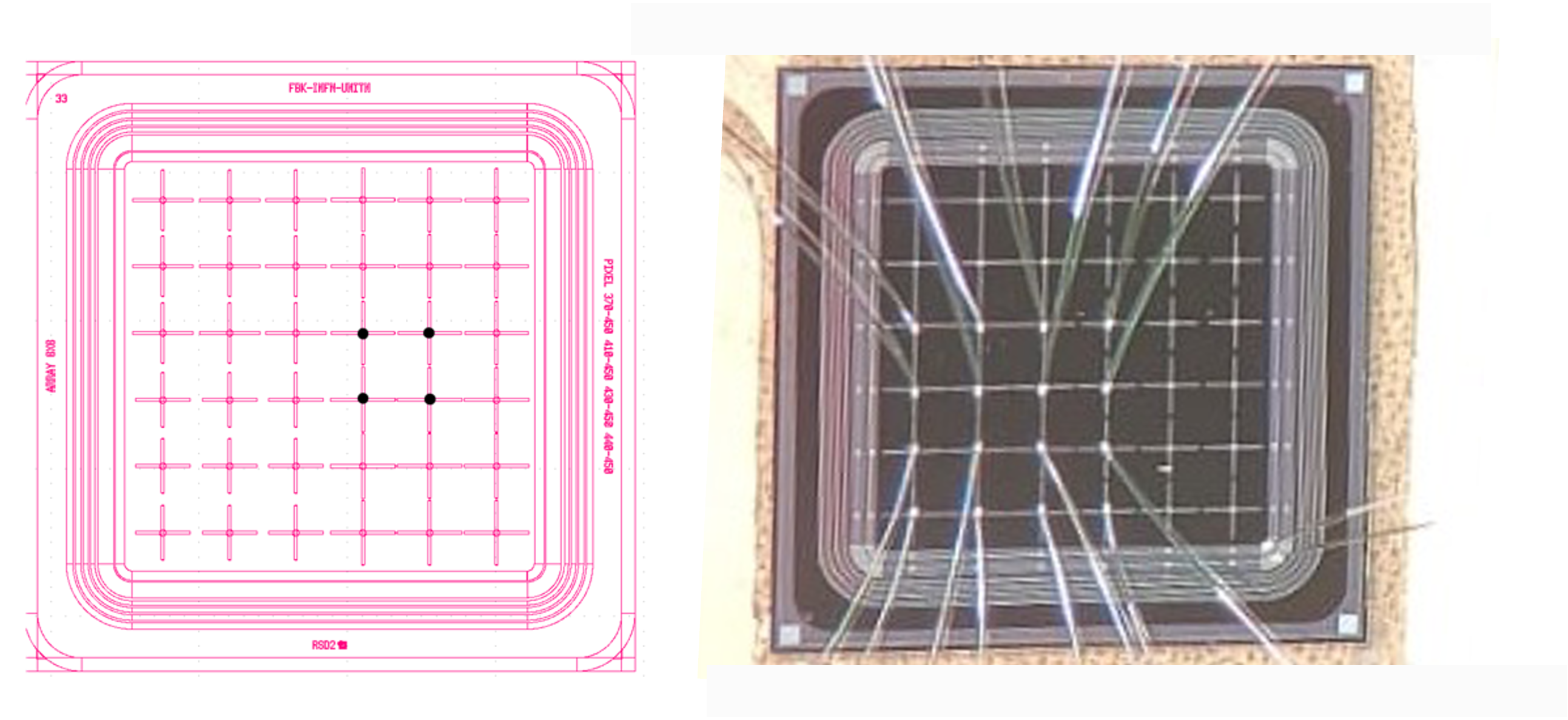}
\caption{Layout and a photograph of a sensor from the RSD2 FBK production with cross-shaped electrodes. } 
\label{fig:RSD2}
\end{center}
\end{figure}

The second RSD production at FBK~\cite{Mandurrino:2021cuq} (RSD2) was designed to test several optimizations of the first FBK RSD production. An in-depth study of the RSD2 spatial and temporal resolutions can be found in~\cite{4DArci}.  An important improvement has been the introduction of cross-shaped electrodes, which significantly increase the response uniformity. Figure~\ref{fig:RSD2} shows on the left pane the layout of a sensor composed of an array of 450 \microns pixels with cross-shaped electrodes while on the right, a photograph of the sensors. Wirebonders are visible for a group of 16 electrodes. 

\begin{figure}[htb]
\begin{center}
\includegraphics[width=0.8\textwidth]{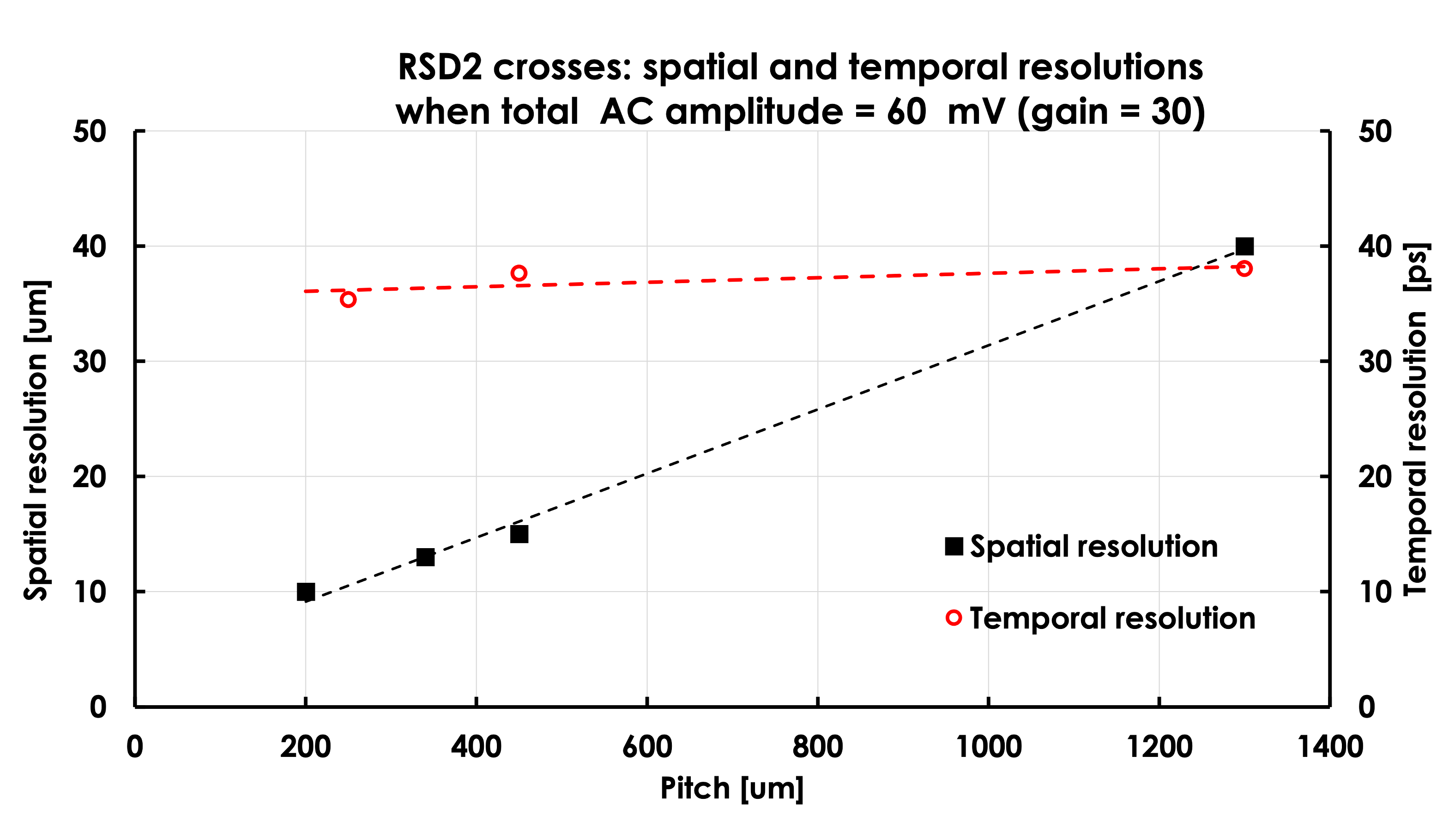}
\caption{Summary of the spatial and temporal resolutions as a function of the pixel size for RSD2 sensors with cross-shaped electrodes~\cite{4DArci}.} 
\label{fig:RSD2Sum}
\end{center}
\end{figure}

The combined spatial and temporal resolutions for several sensors with cross-shaped electrodes are presented in Figure~\ref{fig:RSD2Sum}. The presented results are obtained at a gain of about 30. 

The two key points are: (i) The spatial resolution is about 3\% of the pixel size, and it scales linearly with the pixel size, (i) the temporal resolution is fairly constant at about 35-40 ps as a function of the pixel size.

\section{Conclusions}
The development of resistive read-out in thin sensors with internal multiplication is driven by the needs of future experiments and offers a viable solution to meet the demanding requirements of future charged particle trackers. More generally, it will be relevant for all applications requiring accurate photons and charged particle localization.  Our preliminary results show that a concurrent spatial resolution of about 3\% of the pixel size and temporal resolution of about 35 - 40 ps is achievable. The development of the read-out electronics should be tailored to the need of the specific application: at low power consumption, a simple ToT system is a good choice, while for higher power consumption, a double ToT read-out or a waveform sampler can be considered. 

\section{Acknowledgments}

We kindly acknowledge the following funding agencies and collaborations: INFN – FBK agreement on sensor production; Dipartimenti di Eccellenza, Univ. of Torino (ex L. 232/2016, art. 1, cc. 314-337); Ministero della Ricerca, Italia, PRIN 2017, Grant 2017L2XKTJ – 4DinSiDe; Ministero della Ricerca, Italia, FARE, Grant R165xr8frt\_fare, Compagnia di San Paolo, Italia, Grant TRAPEZIO 2021.

\bibliography{NC_bibfile_2022_RSD2}

\end{document}